\documentclass[journal=jpccck]{achemso}

\usepackage[latin1]{inputenc}

\usepackage{graphicx}
\usepackage{color}
\usepackage{amsmath}
\usepackage{graphicx}
\usepackage{xcolor}
\usepackage{amsfonts}
\usepackage{amssymb}
\usepackage{soul}

\newcommand{\onlinecite}[1]{\hspace{-1 ex} \nocite{#1}\citenum{#1}}

\providecommand{\U}[1]{\protect\rule{.1in}{.1in}}

\title{
Is the largest aqueous gold cluster a superatom complex? Electronic structure \& optical response of the structurally determined Au$_{146}$(pMBA)$_{57}$.
}

\author{X\'ochitl L\'opez-Lozano}
\author{G. Plascencia-Villa}
\affiliation[UTSA]
{Department of Physics \& Astronomy, The University of Texas at San Antonio, One UTSA circle, 78249-0697 San Antonio, TX., USA}
\author{G. Calero}
\affiliation[UPittsburgh]
{Dpt.\ of Structural Biology, University of Pittsburg, Pittsburg, PA, USA.}
\author{R.L.~Whetten}
\affiliation[UTSA]
{Department of Physics \& Astronomy, The University of Texas at San Antonio, One UTSA circle, 78249-0697 San Antonio, TX., USA}
\author{H.-Ch.~Weissker}
\email{weissker@cinam.univ-mrs.fr}
\affiliation[CINAM]
{Aix Marseille University, CNRS, CINaM UMR 7325, 13288, Marseille, France}
\alsoaffiliation[ETSF]
{European Theoretical Spectroscopy Facility}

\begin{document}

\begin{abstract}

The new water-soluble gold cluster Au$_{146}$(pMBA)$_{57}$, the structure of which has been recently determined at sub-atomic resolution by Vergara \textit{et al.},\cite{vergara} is the largest aqueous gold cluster ever structurally determined and likewise the smallest cluster with a stacking fault. The core presents a twinned truncated octahedron, while additional peripheral gold atoms follow a C$_2$ rotational symmetry.  According to the usual counting rules of the superatom complex (SAC) model, the compound attains a number of 92 SAC electrons if the overall net charge is 3- (three additional electrons). As this is the number of electrons required for a major shell closing, the question arises if Au$_{146}$(pMBA)$_{57}$ should be regarded as a superatom complex. Starting from the experimental coordinates we have analyzed the structure using density-functional theory. The optimized (relaxed) structure retains all the connectivity of the experimental coordinates, while removing much of its irregularities in interatomic distances, thereby enhancing the C2-symmetry feature. Analyzing the angular-momentum projected states, we show that, despite a small gap, the electronic structure does not exhibit SAC model character. In addition,  optical absorption spectra are found to be relatively smooth compared to the example of the Au$_{144}$(SR)$_{60}$ cluster. The Au$_{146}$(SR)$_{57}$ cluster does not derive its stability from SAC character; it cannot be considered a superatom complex.

\end{abstract}


\section{Introduction}

Monolayer-protected gold clusters have been commercially available since many years (``PeptideGold''\cite{hainfeld-99}), and they are used in a multitude of biological and medical investigations over the past decade which all require aqueous-phase compatibility; applications include bioconjugation chemistry,\cite{ackerson-biocon-10} protein tagging,\cite{hainfeld-99} biomolecule labeling,\cite{ackerson-10} inhibition of HIV fusion,\cite{bowman-10} as well as growth inhibition of bacteria.\cite{bresee-11}

But their approved use is severely inhibited by a lack of precise knowledge of their structure, composition, and related fundamental properties. Herein, we present a comprehensive study of the electronic and optical properties of the novel water-soluble gold cluster Au$_{146}$(pMBA)$_{57}$, the structure of which has been recently determined at sub-atomic resolution by Vergara \textit{et al.}\cite{vergara}

The stability and chemical nature of many metal clusters has been found to follow the super-atom model,\cite{khanna-92} in which the delocalized states (derived from the $N$ $s$ electrons in noble metals, where $N$ is the number of noble metal atoms)  belong to distinct values of angular momentum and follow an \textit{Aufbauprinzip} similar to that found for the spherical jellium model.\cite{ekardt-SJBM-83} In medium-sized gold clusters, the $n^*$ delocalized orbitals derive mainly from atomic 6s orbitals. The \textit{Aufbau} rule is\cite{walter-08} \ $|1S^2|1P^6|1D^{10}|2S^21F^{14}|2P^61G^{18}|2D^{10}3S^21H^{22}|...$, leading to shell closures at $n^* = 2, 8, 18, 34, 58, 92, 138...$ which explain particular stability when the electron count matches one of these ``magic numbers.''

This observation was generalized as ``super-atom complex model'' (SAC model) \cite{wyrwas-whetten-07,walter-08} to describe monolayer-protected clusters of roughly globular shape. It assumes that stabilizing ligands may either withdraw electrons or localize electrons into covalent bonds. Assuming the withdrawal of one electron for each of the $M$ pseudohalide ligands and noting that each of the N gold atoms provides one 6s electron, one obtains the requirement that $n^* = N - M -z$ be equal to one of the magic numbers listed above for an electronically closed superatom complex, where z is the overall charge\cite{walter-08}. If electronic shell structure is an important factor in the stability of the compound, there should be a clear gap after the shell closure, which implies a deep minimum in the electronic density of states.\cite{walter-08} Projection of the Kohn-Sham orbitals onto spherical harmonics allows for the decomposition of the electronic density of states,\cite{yoon-07,walter-08} clearly showing the SAC character of  many clusters. 
\\


The number of clusters that have been synthesized, purified, and studied has increased immensely over the last few years.\cite{jin-review-16}  Among the monolayer-covered Au clusters, the size range of approximately 140 to 150 gold atoms is of great interest, with the particularly stable, ubiquitous structure Au$_{144}$(SR)$_{60}$ playing a special role for multiple reasons:

\noindent
(i) Its size is such that the localized surface-plasmon resonance (LSPR) is not yet visible, although an incipient LSPR has been discussed.\cite{malola-13,qian-12}

\noindent
(ii) The cluster has been shown in 2014 to exhibit a multitude of discrete peaks in the absorption spectra which reflect the discrete electronic structure.\cite{weissker-natcomm}

\noindent
(iii) Various groups have synthesized and studied the compound, with different thiolate ligand rest groups, demonstrating a high degree of reproducibility (see Ref.\ \onlinecite{weissker-Au144-structure} for a collection).

\noindent
(iv) Different Au$_{144}$-based alloy clusters have been studied both experimentally and theoretically. In particular, the general effect of alloying with silver was studied,\cite{kumara-11,malola-AuAg144-14,barcaro-14-comment,lopez-13} as well as the interesting case of adding copper which seems to produce a plasmon-like resonance,\cite{dass-AuCu-14,bhattarai-15} the origins of which remain, however, unclear.\cite{sinha-roy-AuCu-17}

\noindent
(v) Moreover, the size of 144 gold atoms is where fcc-based (i.e., bulk-like or similar) structures start to dominate,\cite{negishi-15} whereas for smaller clusters molecule-like structures prevail. Specifically for the size of 144 gold atoms and pMBA ligands, polymorphism has recently been inferred: a truncated octahedron structure may coexist with the well-known icosahedral structure.\cite{jensen-polymorphism-16}

\noindent
(vi) Finally, it has been shown that the electronic structure around the HOMO-LUMO gap of Au$_{144}$(SR)$_{60}$ corresponds to super-atom-like states.\cite{lopez-acevedo-09}
\\

The new monolayer-protected gold cluster Au$_{146}$(pMBA)$_{57}$, the structure of which has been recently determined by Vergara \textit{et al.},\cite{vergara}  is the largest aqueous gold cluster ever structurally determined and likewise the smallest cluster with a stacking fault. The core presents a twinned truncated octahedron, while additional peripheral gold atoms follow a C$_2$ rotational symmetry. With the usual electron counting rules as mentioned above,\cite{walter-08}  a charge state of 3- (three additional electrons) leads to a number of SAC electrons of $146-57-(-3)=92$, which would correspond to a SAC shell closing. This raises the question if such a shell closing could explain the stability of the cluster compound. 

In the present manuscript, we use density-functional theory (DFT) calculations to show that despite the proximity of the magic SAC number of 92 electrons, the  Au$_{146}$(pMBA)$_{57}$ cluster does not derive its sstability from super-atom complex character. After a description of the technical details, we discuss briefly the types of bonds and the bond lengths in the relaxed structure. Subsequently, we analyze the electronic structure and the optical response of the cluster and show that the nature of a superatom complex is not realized.

\section{Technical details}

We start from the experimental structure of Au$_{146}$(pMBA)$_{57}$ determined as reported by Vergara \textit{et al.}\cite{vergara} In order to obtain a tractable size for the \textit{ab initio} calculations, we have removed the restgroup R of all the p-MBA ligands, leaving only the carbon atoms directly bound to the sulfur atoms in place. Subsequently, the structure was saturated by hydrogen atoms (resulting in R = CH$_3$ methyl groups) in such a way that one of the hydrogen atoms pointed to the midpoint of the line connecting the two adjacent Au atoms (i.e., the plane defined by S-C-H intersecting this line at its midpoint). For comparison, we have also studied an isoelectronic-substitution model where each complete ligand molecule (pMBA) is replaced by one Cl atom.

The structure was then relaxed using the VASP code\cite{kresse-93,kresse-furth-96,kresse-joubert-99} using density-functional theory (DFT) with the projector-augmented wave method (PAW)\cite{kresse-joubert-99} and employing the LDA exchange-correlation functional and, for comparison, the  GGA functional PBE,\cite{PBE-GGA-96} until all forces were smaller than 0.06\,eV/\AA.  
A charge state of 3- (three additional electrons) was used to comply with the expected electronic shell closing at 2350 electrons according to the SAC model. Relaxation was slow, it appears clear that the total energy surface is composed of many shallow local minima (in particular, sensitive to the orientation of the methyl-group ligands). The chlorine-substituted model relaxed much faster to very small forces (smaller than 0.001\,eV/\AA). The energy difference between the initial and the relaxed structure is about 17\,eV for the relaxation using the LDA functional.
\\

The optical absorption spectra are calculated by TDDFT using the real-space code {\tt octopus}\cite{octopus,octopus2} with the time-evolution formalism\cite{yabana-bertsch-95} and the PBE exchange-correlation potential\cite{PBE-GGA-96} for coherence with previous studies.\cite{weissker-natcomm,weissker-Au144-structure} A comparison of LDA and PBE spectra for identical geometries is shown in Fig.\ S4 the ESI. Norm-conserving Troullier-Martins pseudopotentials\cite{troullier-martins} have been used which, similarly as in our previous work,\cite{weissker-natcomm,weissker-Au144-structure} include the $d$ electrons in the valence (11 valence electrons for each Au atom, i.e., in total 2,350 electrons for the R=CH$_3$ calculations). The spacing of the real-space grid was set to 0.20~\AA, the radius of the spheres centered around each atom which make up the calculation domain to 5~\AA. 

The total energy is used to monitor the stability of the propagation. The time step for the propagation was set to 0.00197\,fs, the propagation time was 25\,fs, and the propagation was carried out by means of the Approximated Enforced Time-Reversal Symmetry (AETRS) propagator\cite{propagators} as implemented in the {\tt octopus} code. The 25-fs period provides a transform-limited resolution of 40 THz, or $\sim$ 0.16 eV, for all the spectra computed and shown in the Figures reported herein.

The results of the time-evolution formalism are equivalent to those obtained using the transition-based Casida formalism.\cite{casida-95,casida-96} A comparison of the time-evolution results with a transition-based calculation is shown in the Supplementary Material of Ref.\ \onlinecite{weissker-natcomm} for the thiolate-ligand covered Au$_{38}$ cluster. For a bare Au cluster, a similar comparison is shown in the supplementary material of Ref.\ \onlinecite{lopez2-13}. The good agreement shows likewise that the technical parameters of our calculations are well controlled.

The electronic energies shown herein are from the VASP calculation. The plotted wave functions are from an octopus calculation.  In order to obtain the angular-momentum-projected density of states (PDOS), we use the (Kohn-Sham LDA) wavefunctions from the octopus ground state calculation and project them onto spherical harmonics according to the prescription of reference \onlinecite{walter-08} used for instance in Ref. \onlinecite{lopez-acevedo-09}. Details are given in the ESI.

\section{Results and Discussion}

\begin{figure}[t]
\centering

\includegraphics[width= .5\columnwidth]{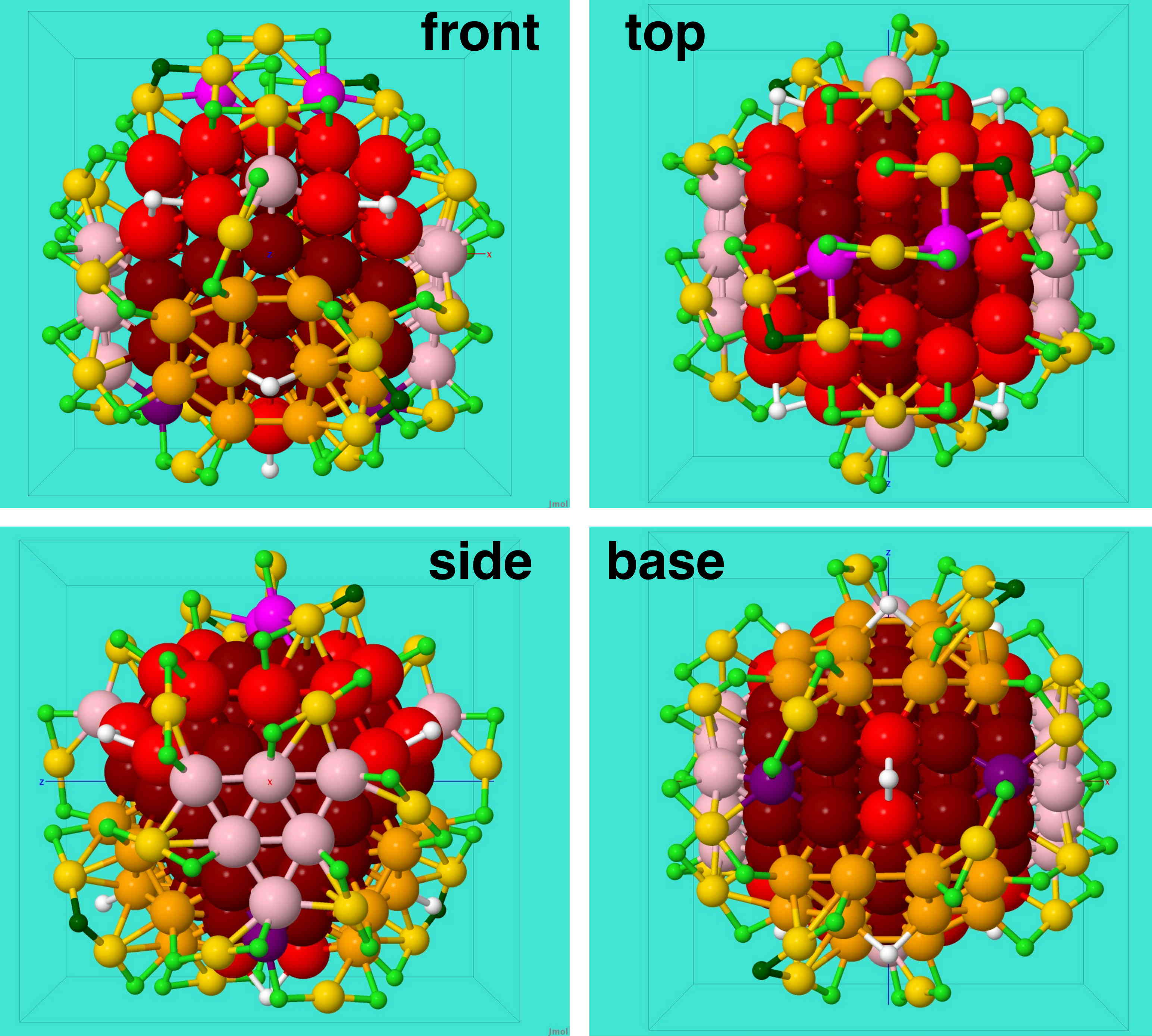}

\vspace{.5cm}

\caption{\textbf{Geometry of the Au$_{146}$(SR)$_{57}$ cluster compound} with characteristic sites highlighted. Shown are the gold sites colored as described in the following, and the S atoms of the ligands. A complete figure of Au$_{146}$(SCH$_3$)$_{57}$ is shown below in Fig. \ref{fig:distances} and in Fig. S2 of the ESI. The cluster holds C$_2$ symmetry (the vertical y-axis corresponds to the two-fold rotation axis). Consequently, 4 of the possible 6 views suffice to present the cluster fully. ``Top'' and ``base'' views are parallel to this two-fold axis.
\newline \hspace{\textwidth}
\textbf{Color scheme for the 146 Au-sites:}
No Au-S bond -- dark red (59); 
single Au-S 'surface' sites (60); 
red (20+2), to complete the twinned truncated-octahedron 81-site core;
pink (14), magenta (2), purple (2), also on FCC lattice (99)
orange (10+10), completes sum to 119. (A rendition of this is seen in the ESI.)
Apart from this core, there are 27 peripheral gold sites shown in yellow (small) in S-Au-S collinear bonds.
\newline \hspace{\textwidth}
\textbf{Color scheme for 57 S-sites:}
7 sites bridging two core-Au-sites -- white;
4 sites bridging two peripheral-Au-sites -- dark green;
38 + 8 = 46 sites bridging surface and core sites -- lime-green.
\label{fig:structure}
}
\end{figure}

In Fig.\ \ref{fig:structure} we show the structure of the Au$_{146}$(SR)$_{57}$ cluster compound with different sites color-coded. All gold sites and the S sites of the ligands are shown. The cluster is approximately globular and exhibits C$_2$ symmetry. (The top view looks down along the two-fold rotation axis.) It's core consists of a twinned truncated octahedron, which is surrounded by an ordered layer of peripheral gold atoms and p-MBA ligands.

\begin{figure}[p]
\centering

\includegraphics[width= .5\columnwidth]{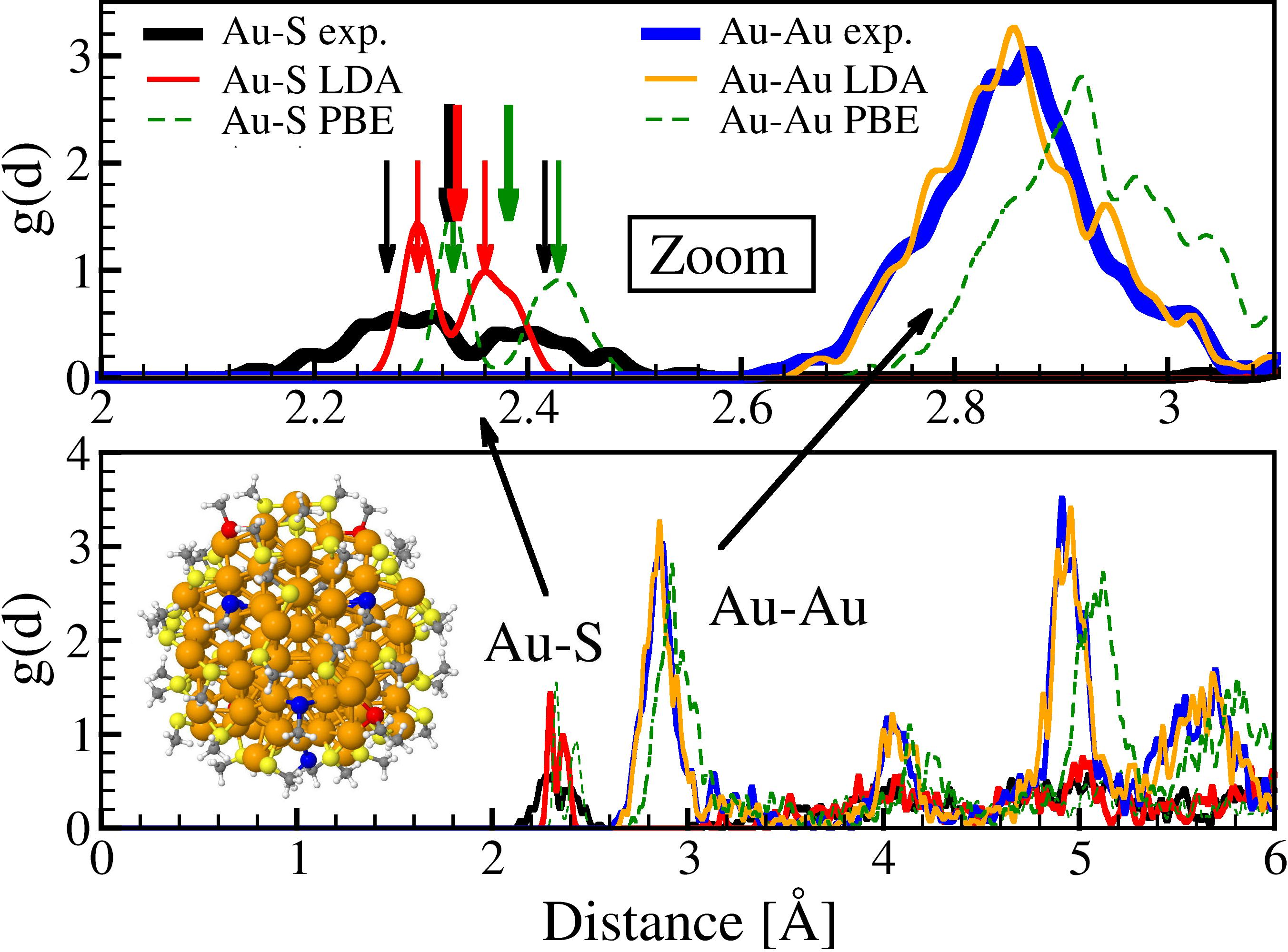}

\vspace{.5cm}

\caption{\textbf{Interatomic distances:} We show the distribution $g(d)$ of Au-Au and Au-S interatomic distances over a wider range (lower panel), with the upper panel showing a blow-up of the region of Au-Au and Au-S \textit{bonds.} 
\newline \hspace{\textwidth}
We compare the unrelaxed structure from experiment \textit{as is} (thick blue and black) with the structure relaxed using the LDA functional and in the 3- charge state. For comparison, the result obtained using PBE-GGA is also shown (dashed green). The thin arrows indicate the average (weight) value of the peaks, using the same color as the curves. The first peak of the experiment is centered at 2.268\,\AA, the second at 2.416\,\AA. The narrower peaks of the LDA-relaxed structure are centered at 2.297\,\AA\ and 2.360\.\AA. Interestingly, the average of all Au-S bonds (thick arrows) have almost the same distance: 2.327\,\AA\ for the experiment, and 2.334\,\AA\ after relaxation (LDA). By comparison, the PBE relaxation led consistently to larger values.
\newline \hspace{\textwidth}
The inset in the lower panel shows the LDA-relaxed cluster including the methyl groups. To highlight the spatial distribution of the Au-S bond lengths, the S atoms are colored blue for those with two S-Au bonds longer than 2.330\,\AA. These are the sites that connect to two core-Au sites, shown in white in Fig.\ \ref{fig:structure}. Colored red are those with two S-Au bonds shorter than 2.330\,\AA\ (bridging two peripheral sites, dark green in Fig.\ \ref{fig:structure}. (The minimum separating the two peaks of the distribution of Au-S bonds is at 2.330\,\AA\ for the LDA relaxed structure.) The yellow ones have both one bond longer and one shorter than 2.330\,\AA.  This corresponds to the situation in Au$_{144}$(SCH$_3$)$_{60}$ where in the ``staple motifs'' the two collinear S-Au-S bonds are shorter than the S-Au bonds connecting to the core.\cite{weissker-Au144-structure} All curves were obtained by convolution with a Gaussian of $\sigma$ = 0.01\,\AA.
\label{fig:distances}
}
\end{figure}

The distribution of interatomic distances,  $g(d)$, is shown in Fig.\ \ref{fig:distances} where we compare the values found for the experimental structure \textit{as is} with the LDA-relaxed structure and, for comparison, with the structure relaxed using the PBE GGA functional.

Clearly, the Au-Au distances are very well reproduced with LDA, which holds not only for the nearest-neighbor distances but also for the Au-Au distances of more distant Au atoms. By contrast, relaxation using the GGA functional PBE overestimates the bond lengths significantly. This is a direct confirmation of the commonly known fact that LDA yields good interatomic distances for gold, which is \textit{ex post facto} a justification of using it in the present calculations. However, the preference of the LDA functional over PBE is certainly less well justified for the Au-S bonds.

Unlike the nearest-neighbor Au-Au bonds, the Au-S bond lengths show a bi-modal distribution already in the experimental coordinates, where, however, this double-peak structure is rather diffuse. These peaks get much narrower following the relaxation, indicating a slight symmetrization of the structure compared to the ``noisy'' data from the (average of multiple) diffraction experiments. The thin arrows indicate the average (weight) value of the peaks, using the same color as the curves. It is interesting to note that in spite of the differences between the LDA-relaxed and the experimentally determined Au-S bonds, their averages as indicated by the thicker arrows in Fig.\ \ref{fig:distances} are almost identical, which serves as a further justification for the use of LDA in the relaxation.

In order to understand the origin of this bimodal distribution, we turn to the colored representation in the inset of the lower panel of Fig.\ \ref{fig:distances}. We color in red the S atoms which have two bonds shorter than 2.330\,\AA. (This is the value of the minimum separating the two peaks.) It turns out that these are exactly the four S sites that bridge two peripheral Au sites, as colored in dark green in Fig.\ \ref{fig:structure}. By contrast, the S atoms which have two bonds that are longer than  2.330\,\AA, rendered in blue in Fig.\ \ref{fig:distances}, turn out to connect two core Au sites (cf., white in Fig.\ \ref{fig:structure}). All the others (yellow) belong to ``simple bridging thiolates'' with one bond longer and one bond shorter than  2.330\,\AA. Inspection shows that this is similar to the situation found for the Au$_{144}$(SCH$_3$)$_{60}$ cluster where in the so-called ``staple motifs'' the two collinear S-Au-S bonds are shorter than the Au-S bonds connecting to the core.\cite{weissker-Au144-structure}

In conclusion of this part, we have a rather good justification for the use of the LDA functional in the calculation, and the character of the geometric features of the ligand layer is in agreement with previously obtained results of the Au$_{144}$ cluster.
\\

\begin{figure}[p]

\centering

\hspace{-.3cm}
\includegraphics[width= .49\columnwidth]{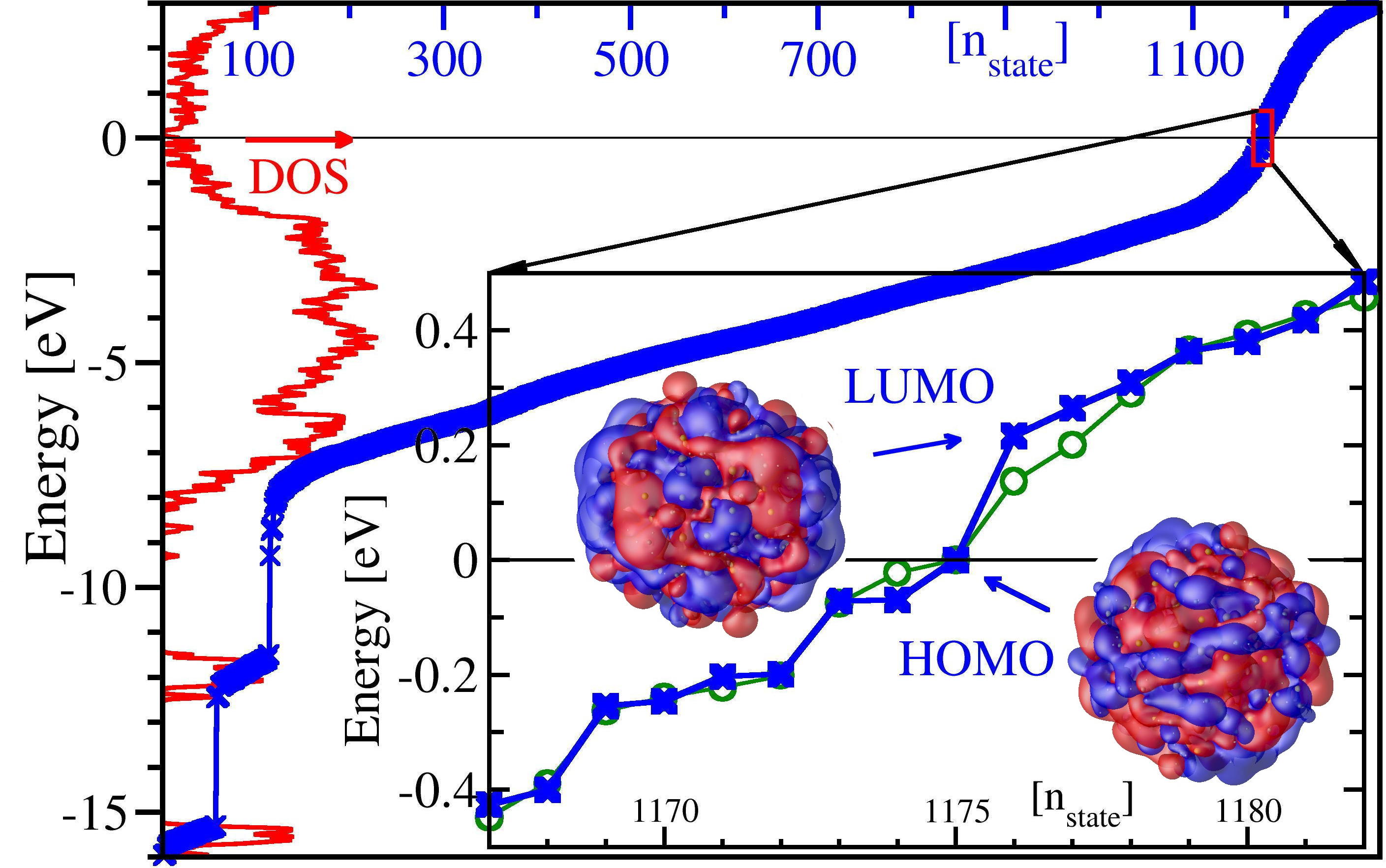}

\vspace{.3cm}

\includegraphics[width= .47\columnwidth]{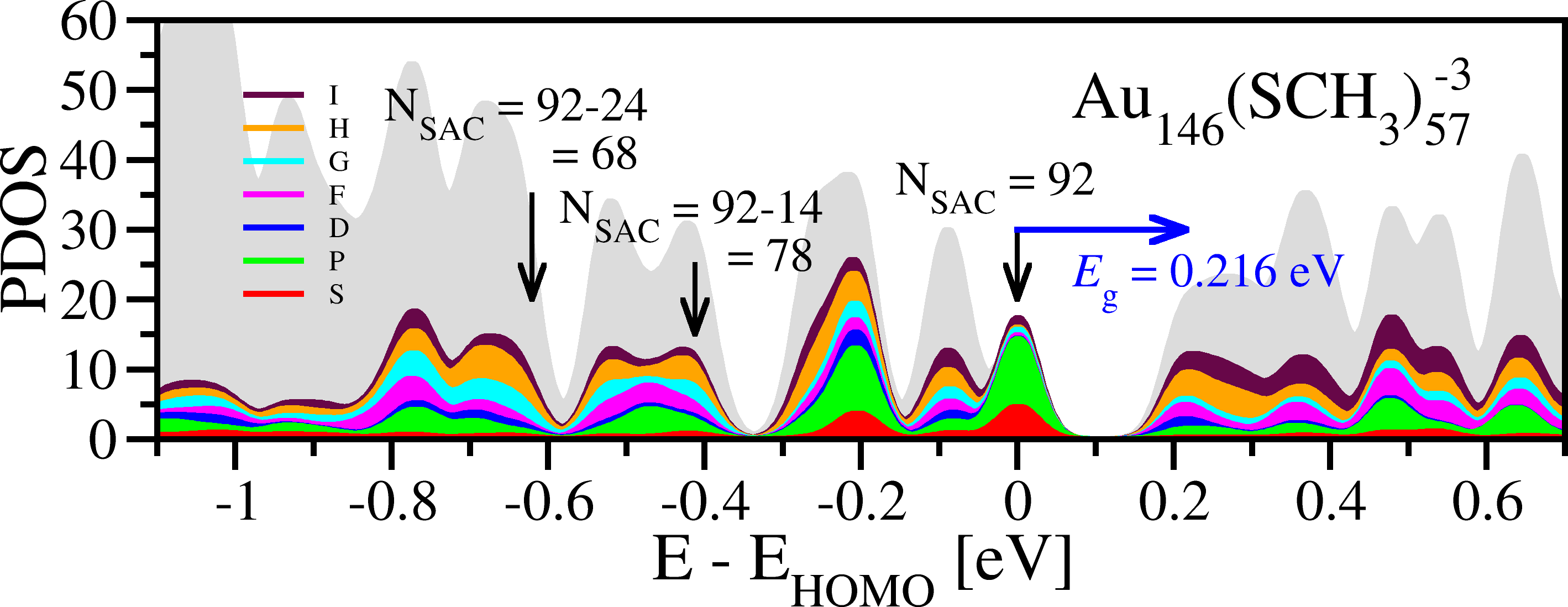}

\includegraphics[width= .47\columnwidth]{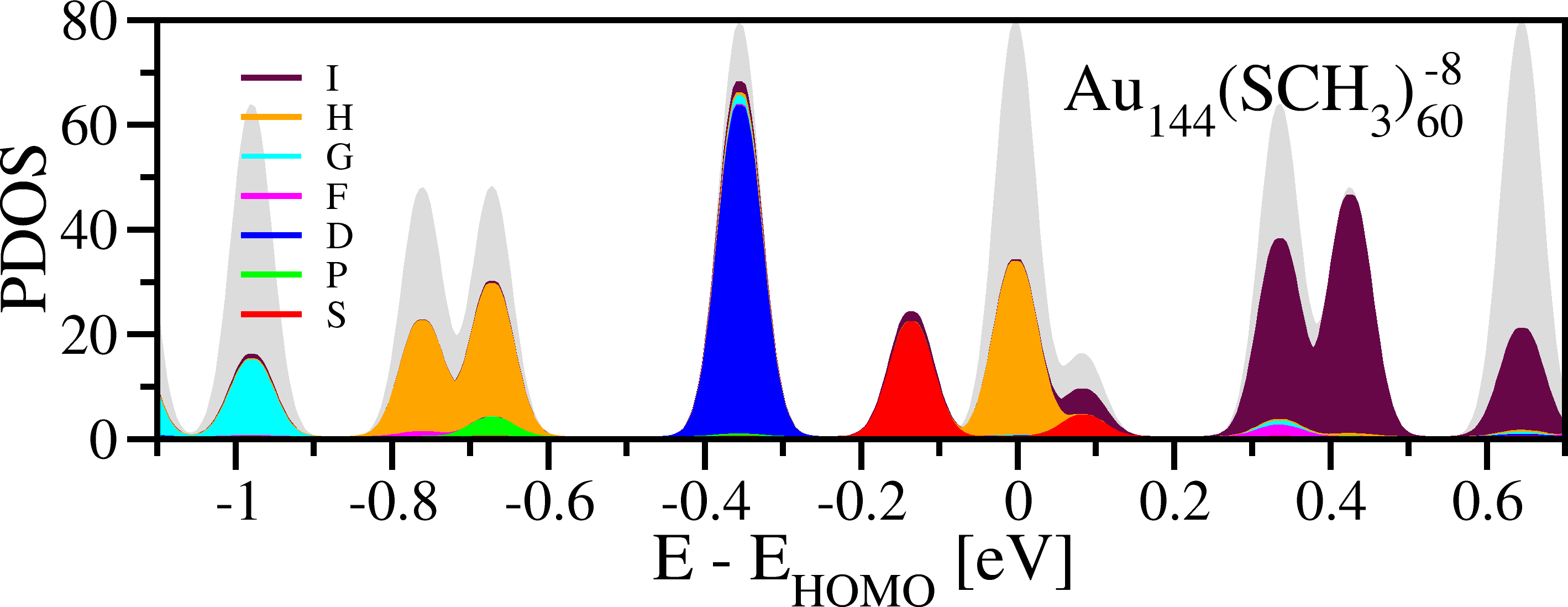}

\vspace{.5cm}

\caption{\textbf{Electronic structure of Au$_{146}$(SCH$_3$)$_{57}^{-3}$.} In the upper panel, we show the electronic density of states (red) as well as the individual energies (blue crosses) as a function of their state number. (The HOMO state corresponds to state number 1175 (2-fold degenerate for spin.)) The inset shows a zoom of the region around the gap for the relaxed structure (blue crosses) and the \textit{as-is} experimental structure (green circles). In addition, the wavefunctions of the nondegenerate HOMO and LUMO states are shown (iso value of 0.001\,\AA$^{-3}$.)
\newline \hspace{\textwidth}
Below, we show the angular-momentum-projected DOS of Au$_{146}$(SCH$_3$)$_{57}^{-3}$. Clearly, in spite of the small gap at the potential shell closing at 92 SAC electrons, the electronic states do not show any clear SAC character, as they are by and large mixtures of different angular-momentum components. This contrasts strongly with the case of the fully symmetric Au$_{144}$(SCH$_3$)$_{60}^{-8}$ cluster, shown for comparison, the states of which exhibit clear SAC character. (The high charge state -8 was chosen to obtain a number of SAC electrons of 92; the structure was from the neutral calculation.) Note that the result is slightly modified compared to the projected DOS of Ref.\ \protect\onlinecite{lopez-acevedo-09} because the fully symmetrized structure of Ref.\ \protect\onlinecite{bahena-13} has been used in the present calculation.  The angular-momentum components are as indicated in the legend, while the grey area reports the full, integrated DOS. Note that the DOS of Au$_{144}$(SCH$_3$)$_{60}$ is more strongly peaked than that of Au$_{146}$(SCH$_3$)$_{57}^{-3}$ due to the much higher symmetry-related degeneracy.
\newline \hspace{\textwidth}
All energies are Kohn-Sham energies from the LDA calculation; the DOS is convoluted with a gaussian of $\sigma$ = 0.025\,eV. The zero of all energy axes was set to the HOMO energy.
\label{fig:dos}
}
\end{figure}

In Fig.\ \ref{fig:dos} we show the results of the electronic structure calculations of Au$_{146}$(SCH$_3$)$_{57}^{-3}$. In the upper panel, we show the full density of states and the individual electronic energies. As usual, one sees the clear Au 5d band and additional peaks originating from the ligands. The region around the HOMO-LUMO gap is enlarged in the inset, with isosurface representations of the frontier orbitals shown. The zero of the energy axis has been set to the HOMO energy.

Comparing the electronic energies of the relaxed cluster in Fig.\ \ref{fig:dos} with those shown for comparison for the \textit{as is}  experimental structure, one observes the effect that was to be expected from the findings of Fig.\ \ref{fig:distances} which shows a symmetrization, i.e., a narrowing of the distribution of the Au-S bonds. While the eigenenergies of the unrelaxed structure are more or less equally distributed, with no discernible differences, neither degeneracies nor an appreciable gap, the gap increases significantly under the relaxation, reaching the value of 0.22\,eV. Moreover, the cumulative density of states shows an inflection point, corresponding to a minimum in the density of states as it is characteristic for a SAC shell closure and indicative of ``electronic stability''.\cite{walter-08} Nonetheless, this gap remains rather small, even in view of the possibility that the LDA underestimates the gap compared to the true value. Thus no clear conclusion about any SAC character can be drawn from the gap value.

For a clearer conclusion concerning the SAC character, the angular-momentum-projected density of states is shown in the second panel of Fig.\ \ref{fig:dos}, again choosing the HOMO energy as origin of the energy scale. It becomes clear here that despite the small gap that might have been indicative of a SAC shell closure, the electronic states are mixtures of many different angular-momentum components. In particular, the strong P contribution (green) would \textit{not} be consistent with the \textit{Aufbau} rule expected in case of a SAC shell closing at 92 electrons. Upon visual inspection of the wavefunctions (Fig.\ S2 of the ESI) we conjecture that maybe a modified ellipsoidal SAC model would be better adapted to describe the cluster. In Fig.\ S2 of the ESI we show the frontier orbitals in different representations for illustration, along with the geometry of the R=CH$_3$ model used in the calculation.

To further confirm these findings, we compare with the case of the fully I$_h$-symmetric Au$_{144}$(SCH$_3$)$_{60}$ cluster\cite{bahena-13,weissker-natcomm,weissker-Au144-structure} in the lower panel of Fig.\ \ref{fig:dos}. (The high charge 8- was chosen to obtain a SAC electron count of 92, corresponding to a possible shell closure. The structure was that of the neutral calculation.\cite{bahena-13}) The difference between the two systems is striking: most of the Au$_{144}$ states exhibit rather pure SAC character with distinct angular-momenta, even though the condition of a shell closing at 92 SAC electrons is not fulfilled. (The HOMO-LUMO gap is actually very small; the SAC gap is at HOMO+1.) Second, the DOS is more peaked in this case, indicating a higher degree of degeneracy due to the high symmetry.
\\

\begin{figure}[p]

\centering

\includegraphics[width= .5\columnwidth]{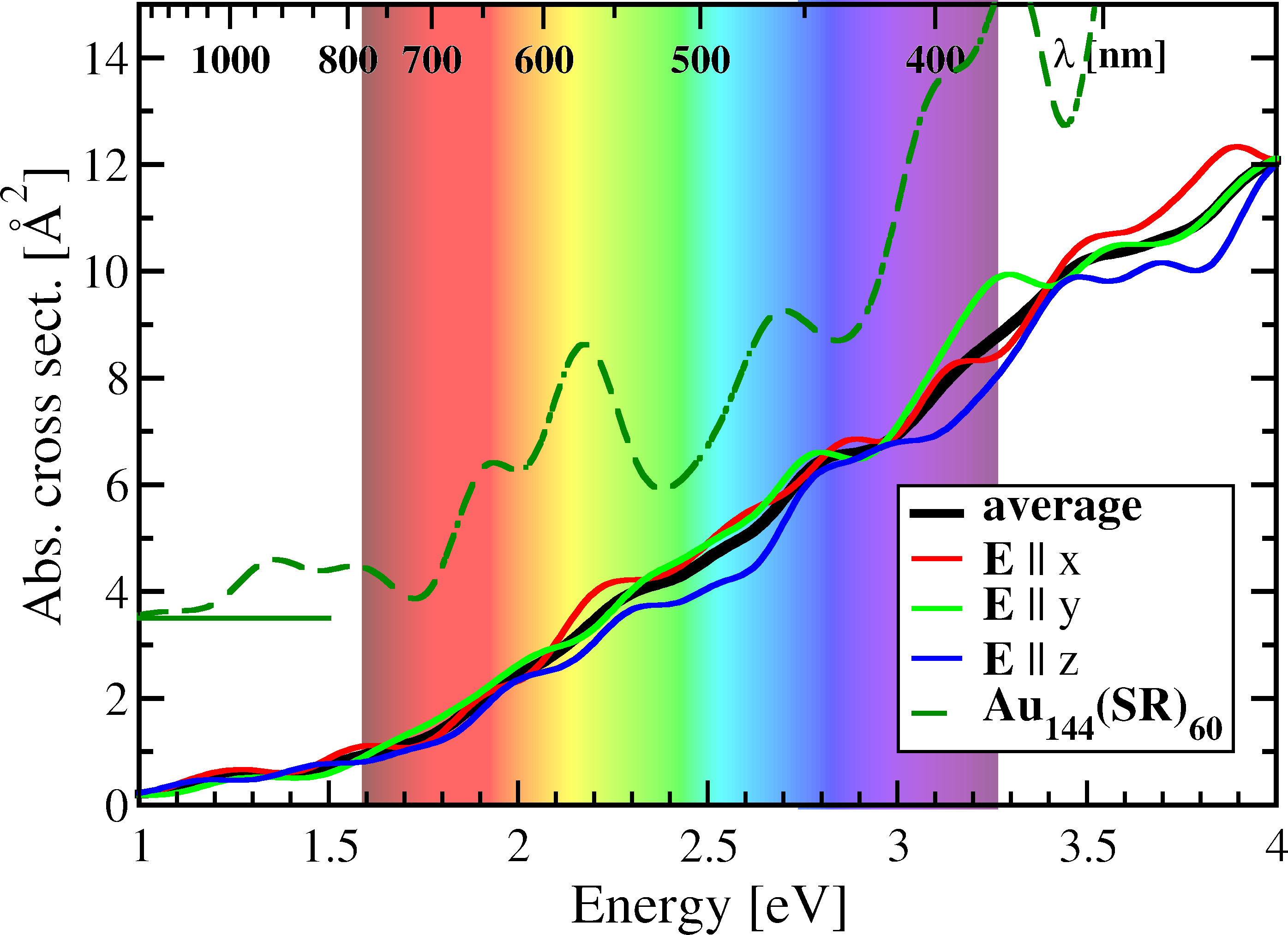}

\vspace{.5cm}

\caption{\textbf{Optical absorption spectrum of Au$_{146}$(SCH$_3$)$_{57}^{-3}$.} The black line shows the average (trace of the full absorption cross-section tensor), while the red, green, and blue lines show the absorption for light polarized along x, y, and z, respectively. Clearly, the system is not isotropic. However, its spectra show comparably little structure compared with the spectrum of the Au$_{144}$(SCH$_3$)$_{60}$ cluster which is shown for comparison (Green dashed line, shifted for visibility). The latter system is isotropic due to its high symmetry (I$_h$). Moreover, this high symmetry leads to higher degeneracy of electronic states, which in turn results in clear, discrete structures in the spectra. The electronic states of Au$_{146}$(SCH$_3$)$_{57}^{-3}$, by contrast, show little degeneracy (cf., Fig.\ \ref{fig:dos}) and the resulting optical response spectrum is rather smooth. (The spectra have been calculated using the PBE functional in order to be consistent with earlier calculations; a comparison with LDA spectra is shown in the ESI.) 
\label{fig:spectra}
}
\end{figure}

In Fig.\ \ref{fig:spectra}, we report the calculated absorption spectra for light polarized along the Cartesian directions alongside the absorption averaged over all directions that would be measured in a solution-phase experiment where the orientation of the clusters is random. As to be expected in view of the symmetry, the absorption of the cluster is not isotropic. However, the differences between the directions are relatively small. 

The average absorption of Au$_{146}$(SCH$_3$)$_{57}$ shows comparably little structure, in particular by comparison with the spectrum of the Au$_{144}$(SCH$_3$)$_{60}$ cluster which is isotropic due to the high symmetry, and which has been reported before.\cite{weissker-natcomm,weissker-Au144-structure} The reasons for this remarkable difference are two-fold: first, the average over the different directions has a smoothing effect. Second, the I$_h$ symmetry of Au$_{144}$(SCH$_3$)$_{60}$ leads to a high degree of degeneracy in the electronic states, which is in turn visible in the spectra. By comparison, as mentioned above, Au$_{146}$(SCH$_3$)$_{57}$ shows little symmetry-related degeneracy; its DOS is less peaked than that of the icosahedral Au$_{144}$ cluster, which translates into  smoother spectra.

These findings mean that the optical spectra measured in solution will not easily find individual spectral structures that calculations could be compared with. We hope that these results motivate the measurement of optical absorption on pure Au$_{146}$(pMBA)$_{57}$ crystals in order to obtain spectra for well-defined directions.

\section{Conclusions}

We have studied the electronic structure and the optical properties of the novel monolayer-protected gold cluster Au$_{146}$(pMBA)$_{57}$, the structure of which has been recently determined by Vergara \textit{et al.}\cite{vergara} It is the largest aqueous gold cluster ever structurally determined and likewise the smallest cluster with a stacking fault.

The density-functional theory calculations were carried out for the charge state 3- (three additional electrons), motivated by the fact that this charge results in a number of 92 SAC electrons for which a major shell closing is expected in case the cluster conforms with the SAC model. The ligand rest groups were replaced by methyl groups, CH$_3$.

Relaxation with the LDA functional resulted in close agreement with the experimental Au-Au distances. The relaxed structure kept all the connectivity of the original coordinates from the experiment, whereas much of the irregularity of the interatomic distances was removed, enhancing the C$_2$ symmetry feature.

Study of the angular-momentum-projected density of states allows us to draw the following central conclusion of the present study: while there is a small gap of the order of 0.2\,eV, the electronic states do not even approximately exhibit SAC character. The strong P-type component of the states close to the HOMO is likewise inconsistant with the \textit{Aufbauprinzip} which would expect even L values to dominate. Thus one cannot attribute the stability of the Au$_{146}$(SR)$_{57}$ compound to a superatomic shell closing, despite the proximity of the magic number of 92 SAC electrons. 

 Finally, we have shown that relatively weak individual structures are found in the absorption spectra. We hope that these results can motivate experimental measurements of the absorption spectra on crystallized samples.

\section{Acknowledgments}

Enlightening discussions with Andres Saul and  Ignacio Garz\'on are gratefully acknowledged. X.L.L. acknowledge previous funding from NSF-DMR-1103730 and NSF-PREM DMR-0934218.  RLW acknowledges the support of the Welch Foundation under grant AX-1857. HCW acknowledges support from the French National Research Agency (Agence Nationale de Recherche, ANR) in the frame of the project ``FIT SPRINGS", ANR-14-CE08-0009. This  work  has  used HPC  resources  from  GENCI-IDRIS (Grant 096829) and from the Laboratory of Computational Nanotechnology and UTSA Research Computing Support Group through the HPC clusters Antares3 and SHAMU, respectively, as well as the Texas Advanced Computing Center (TACC) at the University of Texas at Austin.


\providecommand{\latin}[1]{#1}
\providecommand*\mcitethebibliography{\thebibliography}
\csname @ifundefined\endcsname{endmcitethebibliography}
  {\let\endmcitethebibliography\endthebibliography}{}

\end{document}